\documentclass[]{spie}  

\usepackage{color}
\usepackage{amsmath,amsfonts,amssymb}
\usepackage{graphicx}
\usepackage[colorlinks=true, allcolors=blue]{hyperref}
\usepackage[normalem]{ulem}
\usepackage{marginnote}

\usepackage{hyperref}
\usepackage{todonotes}
\usepackage{textcomp}

\definecolor{mygreen}{rgb}{0,0.5,0}
\definecolor{mygrey}{rgb}{0.5,0.5,0.5}
\definecolor{myred}{rgb}{0.75,0,0}
\definecolor{myblue}{rgb}{0,0,0.75}
\definecolor{mymagenta}{cmyk}{0,1,0,0.12}
\definecolor{mycyan}{cmyk}{1,0,0,0.12}
\definecolor{myorange}{rgb}{1,0.5,0}
\definecolor{myviolet}{rgb}{0.5,0.0,0.75}
\definecolor{mybrown}{cmyk}{0,0.50,1,0.41}

\title{Precision Minimally-destructive detection of ultra-cold atomic ensembles}

\author[a]{Drougakis G.}
\author[a]{Vasilakis G.}
\author[a]{von Klitzing W.}
\affil[a]{Institute of Electronic Structure and Laser, Foundation for Research and Technology - Hellas, P.O. Box 1527, 71110 Heraklion, Greece }

\authorinfo{Further author information: Wolf von Klitzing. E-mail: wvk@iesl.forth.gr}

\begin{document} 
\maketitle

\begin{abstract}
 Over the last two decades the cold-atom physics has matured from proof-of-principle demonstrations to a versatile platform for precision measurements and study of quantum phenomena. Ultra-cold atomic ensembles have been used both for technological and fundamental science applications. To fully exploit their potential, a precise measurement and control of the atom number in the ensemble is crucial. We report on a precise, minimally-destructive measurement technique that can be used to prepare an atomic ensemble with a desired atom number. The measurement relies on the dispersive light-atom interaction, thus it is expected to have a negligible effect on the ensemble temperature and to induce minimal decoherence in the atomic quantum state. As a result, it can be used to perform quantum-enhanced measurements and prepare the atom-number state at the start of an interferometer sequence.   
\end{abstract}


\section{INTRODUCTION}
\label{sec:intro}  

Ultracold atoms have demonstrated great prospects for both technological and fundamental science applications \cite{manifesto,doi:10.1098/rsta.2003.1227}.
To fully exploit their potential, precise control of the atom-number in the cloud is required, without introducing significant disturbance to the atomic quantum state. So far, the majority of the experiments rely on the absorption of near-resonant light by atoms to evaluate the atom number in the ensemble. This leads to significant heating of the atomic clouds and destruction of quantum states of the atoms.
Minimally destructive probing techniques based on the light-atom dispersive interaction have been applied \cite{Deb_2020}. The refractive index of atoms induces a phase shift in off-resonant light; the light phase is then characterized either through interference with a reference beam \cite{Hosten2016} or by applying phase contrast techniques similar to the ones used in microscopy \cite{PhysRevLett.79.553,doi:10.1126/science.273.5271.84}. These techniques can in principle be used to prepare an ultra-cold ensemble at a targeted atom-number size, yet to date their use remains limited. This is mainly due to complexity of experimental apparatus and low signal to noise ratio achieved.
Recently, it was shown that the Faraday paramagnetic effect can be employed to measure the atom number with a resolution better than the atom shot noise for relatively large atomic ensembles \cite{PhysRevLett.117.073604}. Although the technique holds great potential, a rather sophisticated detection system was required for real-time detection.

We report on the development of a robust method for quantifying the number of atoms in a cold atomic ensemble, which can be utilised to prepare ultra-cold ensembles with precisely defined atom numbers.
The measurement is an adaptation of the methodology outlined in \cite{PhysRevLett.117.073604}, employing the Faraday paramagnetic effect as its fundamental principle: off-resonant, linearly-polarized light travelling through a spin-polarised atomic cloud experiences optical rotation at an angle proportional to the number of atoms.
This measurement has a negligible effect on the atomic temperature, so it can be used to prepare the atomic state at the start of an interferometer sequence.
In addition, it holds promise for quantum-enhanced measurements, as demonstrated in the context of warm atoms \cite{PolzikRMP}.

Here, we present a novel technique, which leverages the field-reversal of the Time Orbiting Potential (TOP) magnetic trap \cite{PhysRevLett.74.3352}, a widely employed magnetic trap for ultra-cold-atom experiments. The repeated field-reversal of the TOP allows us to use lock-in detection techniques and moves the Faraday signal from DC to audio frequencies and eliminates or reduces many noise sources. 
Preliminary results of a Faraday signal from an ultra-cold atomic ensemble using this method are presented here. Applications of the proposed research include atomic clocks, inertial sensors, quantum computing, quantum simulations and fundamental physics experiments such as gravitational detectors \cite{El-Neaj2020,Canuel_2020}. 

\section{Methods}
\label{sec:methods}  

The experimental observable is the Stokes polarization component $S_2$  which quantifies the power difference between two orthogonal linear polarizations in the basis ($\pm45^{\circ}$) complementary to the one that the probe-light incident to the ensemble is polarized ($0$ or 90$^{\circ}$). In the limit of large probe detuning from the atomic resonance, the measurement outcome depends  on the Faraday optical rotation angle $\theta_{\text{F}}$ and on the photon flux $\Phi$:

\begin{equation}
 S_2   = \frac{\Phi}{2} \sin(2 \theta_{\text{F}}) \approx  \Phi  \theta_{F} = \Phi \mathcal{G} F_x, \label{eq:chFar:FaradayClassicalS2S1}
\end{equation}
where $\mathcal{G}$ quantifies the light-atom coupling strength \cite{PolzikRMP}, $F_x$ is the total spin component of the atomic ensemble in the direction of probe propagation (taken here to be the $x$ axis), and the approximation holds for small angles. If the trap contains atoms in a well-defined spin state, the ensemble spin component $F_x$ and therefore the Faraday angle is proportional to the atom number: $\theta_F \propto N_{\text{at}}$. We note that magnetic traps are selective to spin states, thus they can be adjusted to confine atoms in a controlled, known Zeeman state. In particular, the TOP trap is formed through the use of a rotating magnetic field, which forces the atomic spin vector to follow the field's rotation. By aligning the TOP trap such that the probe traverses the plane of the rotating field, the Faraday angle signal oscillates synchronously with the field's rotation over time. This provides the advantage of conducting atom number estimation based on a signal frequency away from the low-frequency spectrum susceptible to increased technical noise. 

\begin{figure} [ht]
   \begin{center}
   \begin{tabular}{c} 
   \includegraphics[height=3.5cm]{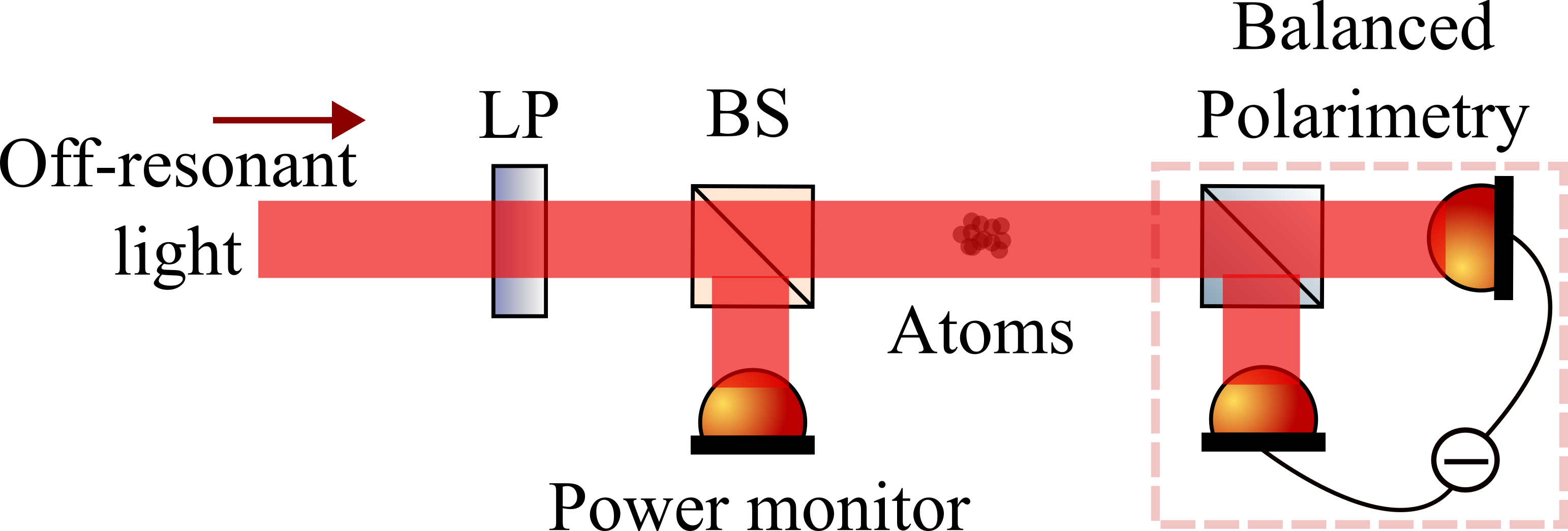}
   \end{tabular}
   \end{center}
   \caption[example] 
   { \label{fig:setup} 
Sketch of the experimental setup used for minimally-destructive detection of atom number in ultra-cold ensembles. }
   \end{figure}

A simplified schematic of the experimental setup is shown in Fig.~\ref{fig:setup}.
The probe light used for the detection was locked a few GHz ($\sim$5~GHz) off-resonant from the atomic transition. A linear polarizer (LP) ensured that the light interacting with the atoms was linearly polarized. A non-polarizing beamsplitter (BS) sent a portion (20\% in power) of the beam to a photodiode to monitor the probe power. As seen from Eq.~\ref{eq:chFar:FaradayClassicalS2S1},  the signal scales with the photon flux; therefore, it is important to know the probe power with a resolution better compared to the targeted atom-number precision.
A fiber collimator delivered a beam with a waist ($\approx 1.3$~mm) at the atomic cloud, which is much larger than the ensemble size (cloud radius $<20$~\textmu m).
Therefore,  the atoms in the ensemble experienced a homogeneous light-intensity distribution to within better than 1\%, rendering trap-position-instability irrelevant for the atom-number characterization.

The $S_2$ observable of light was measured in a balanced polarimetry scheme, where in the absence of any signal the photodiodes measure the same power. A Wollaston prism was used as the analyzing element. The efficiency of light detection was $\approx 87\%$, mainly limited by the photodiode quantum efficiency. The difference between the two photodiode signals was amplified in a transimpedance amplifier and the resulting output was fed into a lock-in amplifier, which demodulated the polarimetry output at the frequency corresponding to the rotation of the magnetic field in the TOP trap.
During an experimental run the data from a lock-in amplifier and the power-monitoring photodiode were recorded.  The polarimetry measurement was normalized with the input power (photodiode signal) to extract the probe-light optical rotation. Care was taken to remove the photodiode offset before the normalization. To reduce the noise in the estimation of the probe power, a moving average technique was implemented, which was sufficient to quantify accurately the slow fluctuations in power.
 
The probed atomic ensemble was $^{87}$Rb, prepared with standard cooling and trapping techniques.
The preparation of the atomic ensemble into a Magneto Optical Trap (MOT) was followed by Compressed MOT (CMOT) and molasses stages, realized with atoms in the $F=2$ hyperfine state.
Then, atoms were transferred with an optical pulse in the absence of repumper light to the $F=1$ hyperfine state and were held in a magnetic quadruple trap; this way, only atoms in the $|F=1, m_F=-1 \rangle $ Zeeman state were trapped.
The atoms were then loaded in to a magnetic quadrupole trap and subsequently cooled by RF-enforced evaporation in the presence of a crossed dipole trap \cite{lin2009rapid}.
The atoms were then loaded to the crossed dipole trap, while maintaining a small quadrupole field displaced by 55~{\textmu}m with respect to the dipole trap in order to prevent spin flips.
After a dipole evaporation ramp for 6~sec, the atoms were loaded in the TOP trap.
This was achieved by adiabatically increasing a homogeneous magnetic field rotating in the horizontal plane (TOP field), while increasing the quadrupole gradient to the targeted value.
After the loading, the most energetic atoms were removed from the trap by applying a strong, RF magnetic field (forced evaporation).

\section{Results}

\begin{figure} [ht!]
   \begin{center}
   \begin{tabular}{c} 
   \includegraphics[height=8cm]{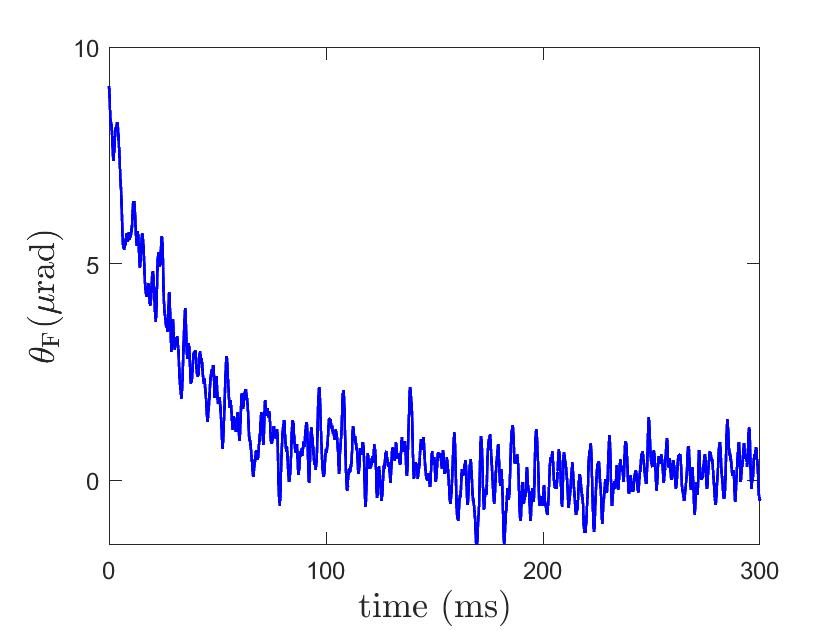}
   \end{tabular}
   \end{center}
   \caption[Decaying atom cloud as observed through the Faraday rotation of the probe beam ] 
   { \label{fig:FaradaySignal} 
Typical demodulated polarimetry output in an experimental run. The observed decay in the signal arises from atom loss primarily attributed to non-zero absorption by atoms of the far-detuned probe light.}
   \end{figure}

A typical demodulated signal of the polarimetry output, normalized to the detected probe power, can be seen in Fig.~\ref{fig:FaradaySignal}.
Each point in the curve represents a measurement of the same trapped atomic ensemble during a single experimental run.
There is a small distortion from the lock-in amplifier, which creates correlations decaying approximately with the lock-in time constant of $\sim$1~ms.
The signal decays due to the atom loss.
The loss-mechanism arises mainly from the non-zero absorption of far-detuned light by atoms, which induces transitions to untrapped atomic states.
The exponentially decaying signal also displays a stochastic behaviour, attributed to photon shot noise \cite{PolzikRMP}, random atom-loss \cite{PhysRevLett.113.263603}, and potentially from experimental imperfections (e.g.\,electronic detection noise).

\section{Conclusions}
In summary, we reported on a method for measuring the atom number in an ultra-cold atomic ensemble. The measurement is based on the Faraday paramagnetic effect and can be readily applied in experiments with cold atoms. It can be used to improve the atom number stability in cold-atom experiments. 

\acknowledgments 
 
The research project was supported by the Hellenic Foundation for Research and Innovation (H.F.R.I.) under the ``2nd Call for H.F.R.I. Research Projects to support Post-Doctoral Researchers'' (Project Number: 768).

\bibliography{report.bib} 
\bibliographystyle{spiebib.bst} 

\end{document}